\documentclass[journal,twoside,web]{IEEEtran}
\usepackage{generic}
\usepackage{amsmath,amssymb,amsfonts}
\usepackage{algorithmic}
\usepackage{graphicx}
\usepackage{textcomp}
\usepackage[style=ieee, locallabelwidth]{biblatex} 
\usepackage{lipsum}
\usepackage{cleveref}
\usepackage{multirow}
\usepackage{array}
\usepackage{siunitx}
\usepackage{xcolor}
\def\BibTeX{{\rm B\kern-.05em{\sc i\kern-.025em b}\kern-.08em
    T\kern-.1667em\lower.7ex\hbox{E}\kern-.125emX}}
{Nima \MakeLowercase{\textit{et al.}}: An Active Dry-Contact Continuous EEG Monitoring System for Seizure Detection Applications in Clinical Neurophysiology}

\title{An Active Dry-Contact Continuous EEG Monitoring System for Seizure Detection Applications in Clinical Neurophysiology}
\author{Nima L. Wickramasinghe$^1$\authorrefmark{1}, Dinuka Sandun Udayantha$^1$, Akila Abeyratne$^1$,  Kavindu Weerasinghe$^1$, Kithmin Wickremasinghe$^2$, Jithangi Wanigasinghe$^3$, Anjula De Silva$^1$, and Chamira U. S. Edussooriya$^1$\authorrefmark{2}
\thanks{$^1$Department of Electronic and Telecommunication Engineering, University of Moratuwa, Sri Lanka}% <-this % stops a space
\thanks{$^2$Department of Electrical and Computer Engineering, University of British Columbia, Canada}%
\thanks{$^3$Department of Pediatrics, Faculty of Medicine, University of Colombo, Sri Lanka}%
\thanks{This work was financially supported in part by the Senate Research Committee, University of Moratuwa.}%
\thanks{\textit{email: \authorrefmark{1}wickramasinghenl.19@uom.lk , \authorrefmark{2}chamira@uom.lk}}
\thanks{\textcopyright  ~2025 IEEE. Personal use of this material is permitted. Permission from IEEE must be obtained for all other uses, in any current or future media, including reprinting/republishing this material for advertising or promotional purposes, creating new collective works, for resale or redistribution to servers or lists, or reuse of any copyrighted component of this work in other works.}
\vspace{-25pt}
}

\begin{document}
\maketitle

\begin{abstract}
% Look at template.tex for original guideline. The abstract must be between 150--250 words. Page limit:8 

%\textit{Methods}: An active electrode is designed with a buffer amplifier and noise reduction techniques to condition the weak EEG signals obtained through dry-contact electrodes. 

Objective: Young children and infants, especially newborns, are highly susceptible to seizures, which, if undetected and untreated, can lead to severe long-term neurological consequences. Early detection typically requires continuous electroencephalography (cEEG) monitoring in hospital settings, involving costly equipment and highly trained specialists. This study presents a low-cost, active dry-contact electrode–based adjustable electroencephalography (EEG) headset, combined with an explainable deep learning model for seizure detection from reduced-montage EEG, and a multimodal artifact removal algorithm to enhance signal quality. Methods: EEG signals were acquired via active electrodes and processed through a custom-designed analog front end for filtering and digitization. The adjustable headset was fabricated using three-dimensional printing and laser cutting to accommodate varying head sizes. The deep learning model was trained to detect neonatal seizures in real time, and a dedicated multimodal algorithm was implemented for artifact removal while preserving seizure-relevant information. System performance was evaluated in a representative clinical setting on a pediatric patient with absence seizures, with simultaneous recordings obtained from the proposed device and a commercial wet-electrode cEEG system for comparison. Results: Signals from the proposed system exhibited a correlation coefficient exceeding 0.8 with those from the commercial device. Signal-to-noise ratio analysis indicated noise mitigation performance comparable to the commercial system. The deep learning model achieved accuracy and recall improvements of 2.76\% and 16.33\%, respectively, over state-of-the-art approaches. The artifact removal algorithm effectively identified and eliminated noise while preserving seizure-related EEG features.

% \textit{Conclusion}: The proposed system addresses key challenges in neonatal seizure diagnosis. Further refinement and clinical trials involving neonates have the potential to revolutionize the early detection and management of neonatal seizures.

\end{abstract}
%\vspace{2ex}

\section{Introduction}

% \IEEEPARstart{E}{lectroencephalography} (EEG) is a critical tool in medical diagnostics and neuroscience, which is used to measure electrical activity in the brain. This technology is important in the medical field for diagnosing and monitoring neurological conditions such as epilepsy, sleep disorders, and brain injuries \cite{teplan2002fundamentals}. In neuroscience, EEG is widely used to study brain function, cognitive processes, and neural dynamics, providing valuable insights into how the brain operates \cite{da2013eeg}. Beyond traditional applications, EEG is also finding new use cases in areas like gaming and neurofeedback \cite{liao2012gaming, badcock2013validation}, where it enables brain-computer interfaces and enhances user experiences through direct brainwave interaction.

\IEEEPARstart{S}{eizures} can occur at any age and result from abnormal neural activity in the brain~\cite{hauser1992seizure}. In newborns, particularly during the first 28~days after birth (the neonatal period), timely detection of seizures is critical to prevent long-term neurodevelopmental impairment~\cite{mizrahi1987characterization,kim2023skin}. However, neonatal seizures are often difficult to recognize, as their clinical manifestations are typically subtle and may be mistaken for normal infant behavior~\cite{mizrahi1987characterization}. Consequently, continuous electroencephalography (cEEG) monitoring in neonatal intensive care units (NICUs) is essential for accurate diagnosis. Globally, the prevalence of neonatal seizures varies with factors such as geographic location, diagnostic criteria, and available diagnostic technologies. Preterm infants are at markedly higher risk, with estimated incidence rates of approximately 95 per 1000 live births, compared to 3 per 1000 live births among full-term infants~\cite{krawiec2023neonatal}. In resource-limited regions, the true incidence may be higher due to underdiagnosis caused by the lack of specialized neurologists and advanced monitoring systems~\cite{wanigasinghe2024seizures}.

In standard clinical practice, electroencephalography (EEG) monitoring is typically performed using wet electrodes to capture brain activity. These electrodes require the application of a conductive gel to reduce skin–electrode impedance and ensure high-quality signal acquisition~\cite{kubota2018continuous}. However, the application and removal of wet electrodes are time-consuming and cause discomfort to patients. Moreover, gel drying during prolonged monitoring can degrade signal quality, often requiring reapplication~\cite{lopez2014dry}.

% \begin{figure*}[t!]
%     \centering
%     \includegraphics[width=1\linewidth]{figures/overall.pdf}
%     \caption{\textbf{Overall block diagram of the system.} The EEG signals captured using the active dry-contact electrodes are passed through to the analog front end for filtering and digitization. The digitized signals are passed to the microcontroller unit, which transmits the data wirelessly through Wi-Fi. The received signals are digitally processed using Python for seizure detection and artifact removal. Processed data is visualized in a Python-based graphical user interface (GUI).
% 1 - Inertial measurement unit sensor,
% 2 - ADS1299 chip,
% 3 - Analog filters,
% 4 - ESP32-S3,
% 5 - Status indicator,
% 6 - Charging indicator.
%  }
%     \label{fig:overall_sys}
%     \vspace{-10pt}
% \end{figure*}

% \begin{figure*}[t!]
%     \centering
%     \includegraphics[width=0.7\linewidth]{figures/overall2.pdf}
%     \caption{\textbf{Overall block diagram of the system.} The EEG signals captured using the active dry-contact electrodes are passed through to the analog front end for filtering and digitization. The digitized signals are passed to the microcontroller unit, which transmits the data wirelessly through Wi-Fi. The received signals are digitally processed using Python for seizure detection and artifact removal. Processed data is visualized in a Python-based graphical user interface (GUI).
%  }
%     \label{fig:overall_sys}
%     \vspace{-10pt}
% \end{figure*}

\begin{figure*}[t]
    \centering
    \includegraphics[width=1\linewidth]{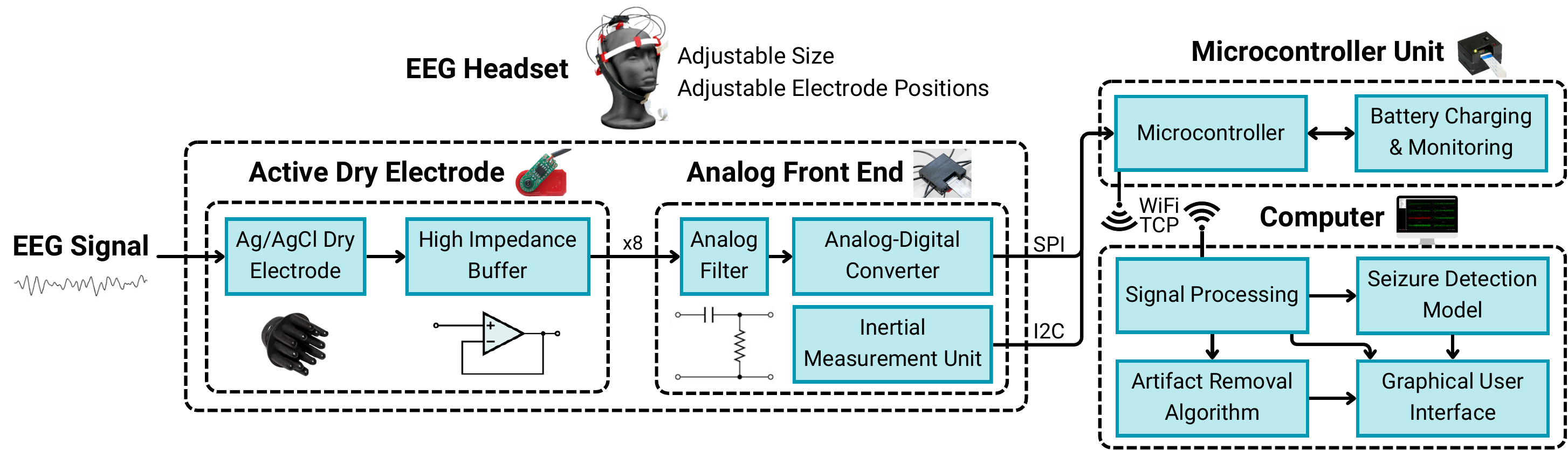}
    \caption{\textbf{Overall block diagram of the system.} The EEG signals acquired by the active dry-contact electrodes are routed to the analog front end (AFE) for filtering and digitization. The digitized signals are then transmitted from the microcontroller unit (MCU) to a host device via a Wi-Fi connection. On the host device, the received data are processed in Python for seizure detection and artifact removal. The processed signals are subsequently visualized using a Python-based graphical user interface (GUI).}
    \label{fig:overall_sys}
    \vspace{-10pt}
\end{figure*}

Although dry-contact electrodes alleviate many of the limitations associated with wet electrodes, they remain highly susceptible to noise and are difficult to maintain in stable contact with the scalp~\cite{chi2010dry,shad2020impedance}. To reduce noise susceptibility, active dry-contact electrodes are employed. These electrodes integrate an operational amplifier (op-amp) buffer with high input impedance positioned in close proximity to the electrode site. The low output impedance of the op-amp buffer strengthens the EEG signal and facilitates noise-resistant transmission~\cite{shad2020impedance}. Previous studies have demonstrated that active dry-contact electrodes can reliably capture EEG signals~\cite{huang2014novel,pourahmad2016evaluation,fonseca2006novel}. 

Because dry-contact electrodes do not inherently adhere well to the scalp, researchers frequently use head caps to improve electrode–scalp contact. For example, \cite{lee2018dry}, \cite{sullivan2008brain}, and \cite{chuang2019cost} have developed custom headset designs incorporating caps for dry-contact electrode applications. In one instance, \cite{lee2018dry} employed a rubber EEG cap to enable simultaneous EEG and functional near-infrared spectroscopy (fNIRS) measurements. However, for clinical use, headset designs must be fully adjustable to accommodate different head sizes and allow electrode placement at clinically relevant positions according to the requirements of the attending clinician.

While EEG monitoring is essential for accurate seizure identification, the shortage of well-trained neurologists in rural hospital settings, coupled with the need for continuous monitoring for reliable diagnosis, has driven the adoption of artificial intelligence (AI) techniques. Reported approaches include support vector machines~\cite{temko2011eeg}, convolutional neural networks (CNNs)~\cite{hossain2019applying}, spatio-temporal networks~\cite{raeisi2023class}, and multilayer perceptron networks combined with Gaussian mixture models~\cite{10675445}. However, many of these methods lack real-time processing capability and model interpretability, both of which are critical for clinical implementation.

% (Should we mention about commercial devices and their problems)

% Existing methods
% \begin{itemize}[noitemsep]
% \item Custom AFEs but not suitable for dry-contact electrodes \cite{chuang2019cost,consul2017neuromonitor,senevirathna2016low,mccrimmon2017performance,acharya2015eeg}
% \item dry-contact electrodes but no AFE
% \cite{huang2014novel,lee2019two,pourahmad2016evaluation}
% \item dry-contact electrodes with AFE but show their problems
% \cite{nathan2015design,gao2018soft,gargiulo2008mobile,lee2018dry,sullivan2008brain}
% \item Commercial devices and their problems
% \cite{hinrichs2020comparison,soufineyestani2020electroencephalography,larocco2020systemic}
% \end{itemize}

Given the significant underdiagnosis of neonatal seizures~\cite{murray2008defining}, primarily due to the inconvenience and high cost of EEG monitoring equipment and the limited availability of specialized neurologists, improving access to affordable monitoring solutions and automated seizure detection technology has the potential to substantially enhance neonatal care. The ability to initiate immediate interventions following the detection of abnormalities can improve survival rates and reduce the risk of long-term neurodevelopmental impairment~\cite{kim2023skin}. To this end, we propose a low-cost, user-friendly cEEG monitoring system with a seizure detection back-end.

The following contributions are made by our work:
\begin{itemize}
    \item an end-to-end active dry-contact electrode-based portable 8-channel EEG acquisition platform made using commercial off-the-shelf components,
    \item a low-cost wearable headset design that is fully adjustable and supports clinical EEG electrode placements, 
    \item a cEEG monitoring system with an integrated real-time interpretable seizure detection model, real-time movement detection, and automatic artifact removal.
\end{itemize}

To assess system performance in a representative clinical setting without the ethical challenges of involving neonates, we conducted validation on an older pediatric patient diagnosed with absence seizures, enabling a safe, controlled demonstration of the device’s ability to capture seizure activity while allowing direct comparison with a commercial wet-electrode EEG system. Quantitative analysis shows a cross-correlation $>$ 80\% between the signals obtained by our custom dry electrode device and the commercial wet electrode EEG system available, with comparable signal-to-noise ratio (SNR) values. The results of the proposed model architecture, initially trained on the Helsinki Zenodo Neonatal EEG dataset~\cite{stevenson2019dataset}, were first reported in our previous work~\cite{AI_paper}, where we presented the training and validation outcomes in detail. In that work, for an 80\%/20\% train-test split, the model achieved absolute improvements of 2.71\% and 16.33\% in area under the curve (AUC) and recall, respectively. Moreover, when evaluated with a 10-fold cross-validation, the model achieved improvements of 8.31\% and 42.86\% in AUC and recall, respectively.

% In \Cref{sec:prop}, we describe the design of our custom active dry-contact electrode EEG acquisition device, the signal processing techniques used, and the development of seizure detection and artifact removal models. In \Cref{sec:results}, we outline the clinical experiment, compare our device's signal quality to a commercial standard wet electrode system, and present the results of seizure detection and artifact removal. Conclusion and future work are presented in \Cref{sec:conc}.

\section{Proposed EEG Acquisition and Seizure Detection System} \label{sec:prop}

The overall block diagram of the system is shown in Fig.~\ref{fig:overall_sys}. In this paper,  \Cref{sec:dry,sec:overall} describe the hardware interface, including the active dry-contact electrodes, the analog front end (AFE), and the microcontroller unit (MCU). \Cref{sec:machine,sec:signal} describe the seizure detection model and the automatic artifact removal algorithm. %The experimental setup used during the clinical experiment is explained in \Cref{sec:experimental}. 
\begin{figure}[t!]
  \centering
  \includegraphics[width=1\linewidth]{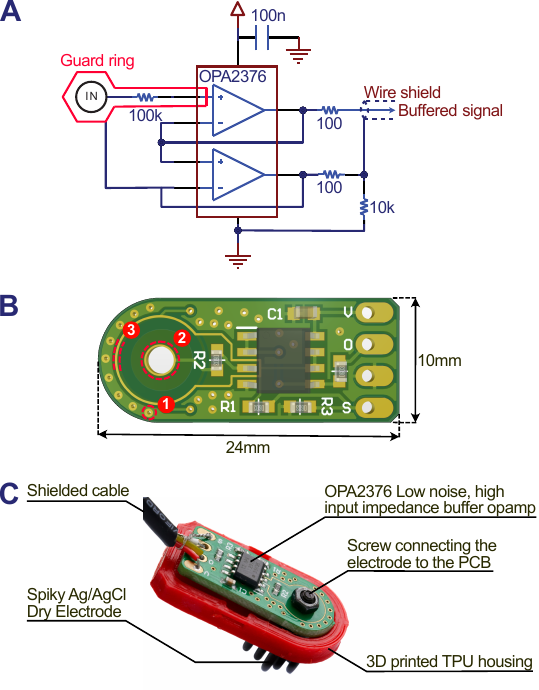}
  \caption {\textbf{An overview of the active dry-contact electrode design. A)} Schematic diagram showing the usage of op-amp buffers. %The first op-amp isolates the weak EEG signal and strengthens it. The output of the first op-amp buffer is fed into the second op-amp buffer, which actively drives the guard ring and the wire shield and isolates the initially buffered signal from any noise interference acting upon the shields. 
  \textbf{B)} PCB layout showing the methods used to minimize noise interference. 1 - Via stitching and via shielding to reduce noise, 2 - Plated through hole and pad to attach the dry-contact electrode using a screw, 3 - High impedance input is actively shielded by the guard ring. 
  \textbf{C)} Assembly of the active electrode on the TPU housing connected to the Spiky Ag/AgCl dry-contact electrode using a screw. }
  \label{fig:active_electrode}
  \vspace{-10pt}
\end{figure}

\subsection{Active Dry-Contact Electrodes and Headset Design}
\label{sec:dry}

The developed active dry-contact electrode, shown in Fig.~\ref{fig:active_electrode}, incorporates impedance buffering directly at the electrode site to enhance EEG signal quality and enable noise-reduced transmission. The design employs Ag/AgCl dry-contact electrodes due to their low skin–electrode impedance~\cite{electrode_material} and relatively low cost. Multi-spike geometries are adopted to improve penetration through hair, ensuring better scalp contact~\cite{electrode_shape}.

The Ag/AgCl electrode is mechanically and electrically coupled to the printed circuit board (PCB) through a conductive pad with an electroless nickel immersion gold (ENIG) surface finish. The PCB is designed with a screw-based mounting mechanism, allowing secure attachment and efficient replacement of electrodes when required. As illustrated in Fig.~\ref{fig:active_electrode}A and Fig.~\ref{fig:active_electrode}B, active shielding is implemented on both the guard ring and the electrode cable~\cite{active_shield, pourahmad2016evaluation}. Additionally, passive shielding techniques, including ground pours and via stitching, are incorporated into the PCB layout to further suppress noise.

The EEG signal is buffered using an OPA2376 operational amplifier (Texas Instruments, Texas, USA), featuring an input bias current of 0.2~pA, a CMOS input stage with typical input impedance in the T$\Omega$ range, low input-referred voltage noise (0.8~\si{\micro\volt_{pp}} in the 0.1--10~Hz range), low input current noise (2~fA/$\sqrt{\text{Hz}}$), and integrated electrostatic discharge (ESD) protection.

% \subsection{Headset Design}
% \label{sec:headset}

\begin{figure}[t]
    \centering
    \includegraphics[width=1\linewidth]{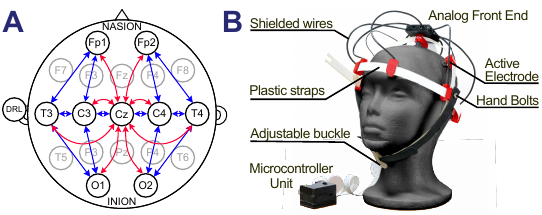}
    \caption{ \textbf{Overall headset design.} \textbf{A) }Reduced electrode montage used in the headset compared to the standard 10-20 EEG electrode system. Red lines show the 8 channels recorded by the hardware. Blue lines show the 12 channels deduced using the 8 channels for the machine learning model and commonly used by medical practitioners. \textbf{B)} Headset design showing the hand bolts used to adjust the length of the plastic strap to fit any head size. The active dry-contact electrodes can be slid on the strap for optimal placement. The adjustable buckle on the bottom fits the headset firmly onto the user and minimizes noise artifacts during movement.}
    \label{fig:headset_design}
    \vspace{-10pt}
\end{figure}

The active dry-contact electrodes are housed within a red enclosure, fabricated via three-dimensional (3D) printing using PLA+ material as shown in Fig.~\ref{fig:active_electrode}C as well as in  Fig.~\ref{fig:headset_design}B. The white strips supporting the active electrodes are made from flexible plastic, enabling both stability and adaptability.

Previous studies have demonstrated that reduced electrode montages can maintain high sensitivity and specificity for neonatal seizure detection compared to the standard 10--20 EEG system~\cite{reduced_eeg, ryan2024electrographic}. Guided by these findings, the headset design as depicted in Fig.~\ref{fig:headset_design}A adopts a 12-channel reduced montage, comprising nine active electrodes and a driven right leg (DRL) electrode for EEG acquisition.
Electrode positions can be adjusted by sliding them along the flexible plastic strips, allowing adaptation to head circumferences ranging from 30~cm to 60~cm, suitable for pediatric patients, and can also be scaled down to below 30~cm for neonates. The horizontal strip encircling the forehead supports the O1, O2, Fp1, and Fp2 electrodes, while the vertical strip spanning the scalp supports the C3, C4, T3, T4, and Cz electrodes. The AFE, positioned at the top of the head, is directly attached to the Cz electrode housing. The strip lengths are adjustable and secured using two hand bolts on the sides. An adjustable-length chin strap with a plastic buckle ensures stable placement for optimal EEG signal acquisition, as shown in Fig.~\ref{fig:headset_design}B.

 \subsection{Analog Front End and Microcontroller Unit}
 \label{sec:overall}

% \begin{figure*}[!ht]
% \centering
% \includegraphics[width=1\linewidth]{figures/SigPipe.pdf}
% \caption{Signal processing pipeline. ML - Machine Learning, PSD - Power Spectral Density, IC - Independent Component, ICA - Independent Component Analysis, GUI - Graphical User Interface}
% \label{fig:sigPipe}
% \vspace{-5pt}
% \end{figure*}

\begin{figure}[t!]
    \centering
    \includegraphics[width=1\linewidth]{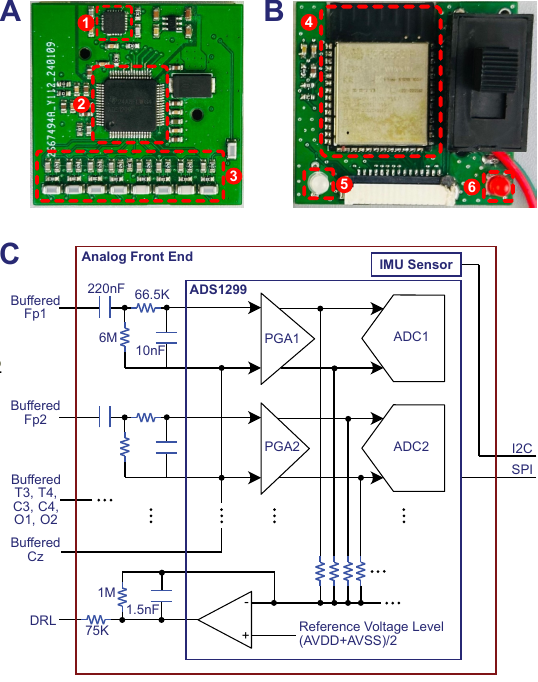}
    \caption{\textbf{Analog front end (AFE) and microcontroller unit (MCU) design.} 
    \textbf{A)} Printed circuit board (PCB) realization of the AFE. 
    \textbf{B)} PCB realization of the MCU.
    1-Inertial measurement unit sensor,
    2-ADS1299 chip,
    3-Analog filters.
    4-ESP32-S3 microcontroller,
    5-Status indicator,
    6-Charging indicator. 
    \textbf{C)} Schematic diagram of the AFE.}
    \label{fig:pcb}
    \vspace{-10pt}
\end{figure}

The buffered signals from the active electrodes are acquired through the AFE, which incorporates the ADS1299 (Texas Instruments, Texas, USA), an 8-channel, 24-bit analog-to-digital converter with low input-referred voltage noise (0.98~\si{\micro\volt_{pp}}), which is shown in Fig.~\ref{fig:pcb}A. This high precision enables noise-minimized voltage measurements down to 0.983~\si{\micro\volt_{pp}} over a large input range of 0.375~V. The device is configured for a sampling rate of 250~Hz, with a gain of 24\si{\times} per channel, and utilizes the internal DRL circuit to enhance common-mode rejection. The ADS1299 configuration for 8-channel signal acquisition and DRL operation is shown in Fig.~\ref{fig:pcb}C. An MPU6050 (TDK InvenSense, USA) 6-axis inertial measurement unit (IMU) is integrated to capture head motion data. Within the AFE, signals first pass through an analog filtering stage comprising two passive resistor–capacitor sections that form a second-order Butterworth bandpass filter. The high-pass edge at 0.1~Hz attenuates slow drifts and baseline wander, preventing amplifier saturation, while the low-pass edge at 120~Hz suppresses high-frequency noise outside the EEG bandwidth (typically above 100~Hz) to reduce aliasing during digitization. Data from the AFE are transmitted to the MCU via a flat flexible cable.

%The AFE utilizes separate ground planes for analog and digital signals connected at a single point. This physical separation minimizes noise coupling between the sensitive EEG signals and the high-frequency digital processing circuitry.
% The AFE was designed to be lightweight to ensure comfort to the user when placed on the head. 

An ESP32-S3 (Espressif Systems, China) shown in Fig.~\ref{fig:pcb}B serves as the primary MCU, providing wireless data transmission from the portable device to a computer. EEG data are transferred over 2.4~GHz Wi-Fi using the transmission control protocol (TCP) to ensure packet integrity and order. The device is powered by a rechargeable 2000~mAh, 3.7--4.2~V lithium–polymer battery, supporting up to 13~h~20~min of continuous operation at a maximum current draw of 150~mA during transmission. An ADUM4160 (Analog Devices, Massachusetts, USA) USB isolator is employed to electrically isolate the device from mains power during charging.

Figure~\ref{fig:pcb} illustrates the fabricated printed circuit boards (PCBs) for the AFE and MCU. The total fabrication cost of the PCBs, electronics, and plastic components for the complete system prototype is approximately 170~USD.

\begin{figure*}[tbp]
% \vspace{-5pt}
    \centering
    \includegraphics[width=1\linewidth]{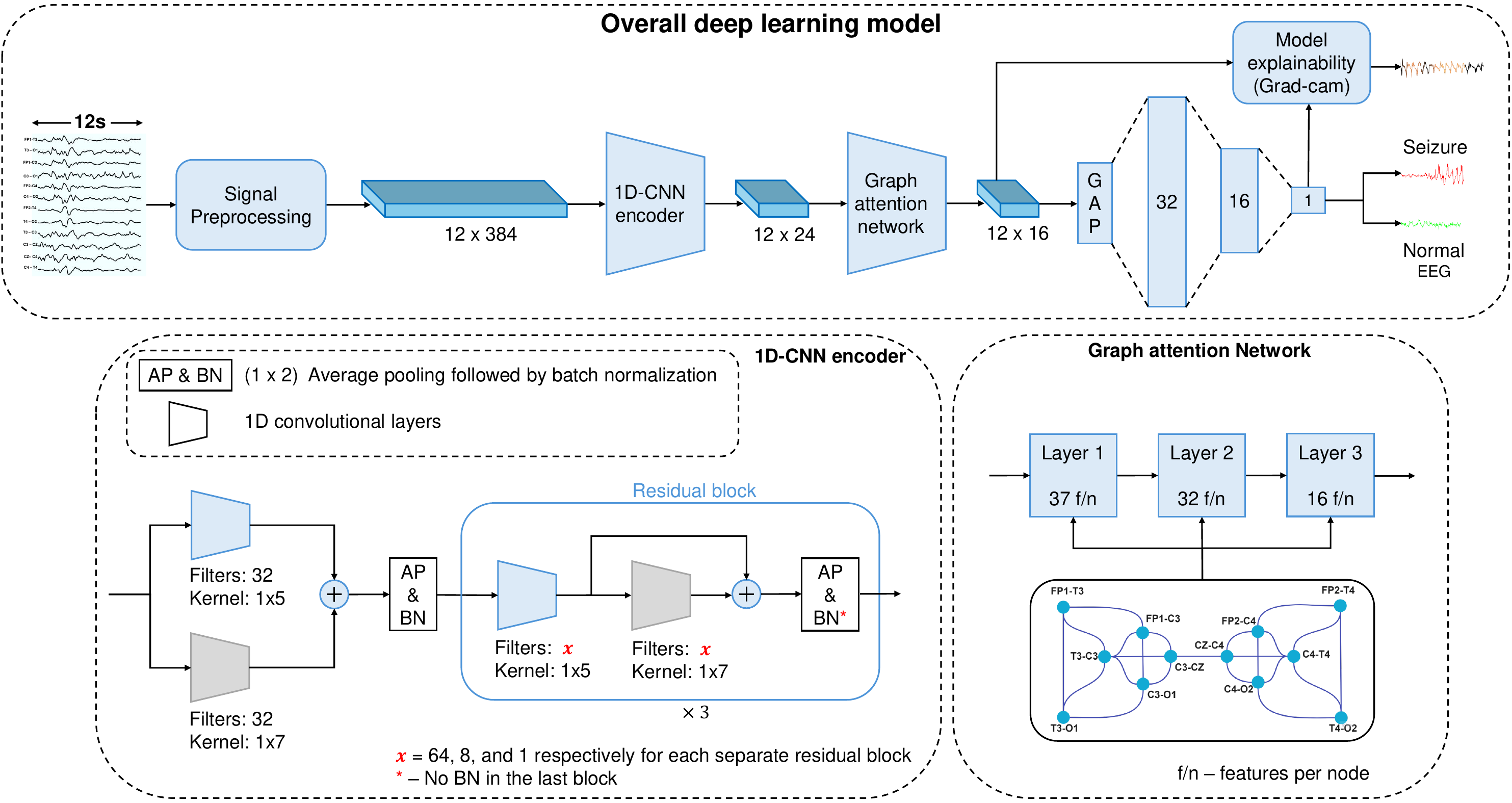}
    \caption{\textbf{Complete overview of deep learning model}. The 12 s long EEG epoch is preprocessed and downsampled to 384 samples for all 12 channels. Then, the 1D-CNN encoder extracts temporal features with either (1$\times$5) or (1$\times$7) convolutions, and the graph attention network extracts spatial features using the given graph. Then, a global average pooling (GAP) layer followed by a fully connected network with 32, 16, and 1 neurons in each dense layer classifies the epoch (binary classification). The grad-cam block explains the model output.}
    \label{fig:deepmodel}
    \vspace{-10pt}
\end{figure*}

\subsection{Signal Pre-processing and Data Preparation}\label{sec:signal}

The acquired analog EEG signals are digitally preprocessed prior to seizure detection to remove high-frequency noise, baseline drift, and powerline interference. These preprocessing steps are performed in real time by the system software. The primary noise sources addressed include baseline drift (0.001--0.5~Hz), powerline noise at 50~Hz and its harmonics, and other high-frequency components above 50~Hz. 

A fourth-order Butterworth bandpass filter with normalized passband edges of 0.004 to 0.4 (relative to the sampling frequency) is implemented to attenuate these noise components, providing a minimum stopband attenuation of 24~dB. In addition, second-order notch filters centered at 50~Hz and 100~Hz, each with a $-$3~dB bandwidth of approximately 4~Hz, are used to further suppress powerline interference.

The 12-channel reduced montage EEG is reconstructed from the 8-channel EEG data acquired by the AFE as shown in Figure \ref{fig:headset_design}A. Data from the IMU is processed using a threshold-based approach to detect head movements and generate alerts.

% The filtered data should be further processed prior to input to the machine learning model. A 12-second window—comprising the 12 seconds immediately preceding and including the current time point—is selected. Then this epoch is passed through a $1-16$Hz bandpass filter followed by downsampling to $32$ samples per second to facilitate real-time ML inference. Therefore, the input to the model is a (12 × 384) signal where 12 is the number of bipolar channels shown by blue arrows in Fig.~\ref{fig:headset_design} and 384 is the number of samples derived by time(12s) × new sampling rate(32Hz).

\subsection{CNN-GAT based Model for Seizure Detection}
\label{sec:machine} 

Inspired by the sequential CNN and graph attention (GAT)-based framework proposed in \cite{raeisi2023class} for neonatal seizure detection using the 18-channel full-montage EEG, a new architecture is developed for the 12-channel reduced-montage EEG. As shown in Fig.~\ref{fig:deepmodel}, the functional CNN encoder consists of four blocks. The first block performs two parallel one-dimensional convolutions with different kernel sizes to extract distinct temporal features, followed by element-wise addition, average pooling, and batch normalization. Following the design principles in \cite{resnet}, the remaining three blocks are implemented as residual blocks to improve performance while minimizing computational complexity. The CNN encoder thus acts as a temporal feature extractor while preserving inter-channel dependencies.

The GAT module models the neonatal brain connectivity within the deep learning framework. In the constructed graph, nodes represent channel-specific features, and edges represent anatomical connections between lobes, derived from the reduced montage configuration. With three GAT layers, each node aggregates information from third-order neighbours, covering approximately 78\% of a well-developed brain’s connectivity. The GAT output is passed through three fully connected layers and a global average pooling layer for binary classification.

The model input size is set to (12$\times$384), where 12 corresponds to the number of EEG channels and 384 to the number of samples in each epoch. Each 12-second epoch—comprising the 12 s immediately preceding and including the current time point—is bandpass-filtered between 1–16~Hz (in addition to the preprocessing described in the previous section) and downsampled to 32~Hz prior to model inference. The resulting architecture contains 46{,}612 learnable and 208 non-learnable parameters, enabling deployment on edge devices without sacrificing real-time performance.

To enhance interpretability, a modified Grad-CAM~\cite{grad-cam} approach is implemented to identify the temporal segments and channels contributing most to the model’s output. By computing the derivatives of the classification logit with respect to the final GAT layer, this method achieves explainability while maintaining real-time processing capability.

\subsection{Hybrid CNN-LSTM Model for Artifact Detection} 

EEG signals are frequently contaminated with artifacts such as muscle activity and eye movements. While wavelet-based artifact removal techniques can operate in real time, they often struggle to distinguish between seizures and artifacts, as certain artifacts (e.g., eye movements) share similar frequency characteristics with seizure patterns. To address this limitation, the proposed approach first applies the extended-infomax independent component analysis (ICA) algorithm~\cite{lee1999independent} to decompose the input EEG into eight independent components. These components are subsequently classified by a machine learning model explicitly designed to label them as either artifacts or non-artifacts, thereby eliminating the need for manual annotation by clinical experts.

The model is trained using the EPIC dataset~\cite{lopes-2022}, which contains ICA-derived components from both seizure and artifact segments. The architecture is inspired by the method of \cite{lopes2022ensemble} and achieves higher artifact-labeling accuracy compared to the widely used MNE-ICAlabel algorithm~\cite{Li2022}. It comprises three deep neural networks (DNNs): (i) a bidirectional long short-term memory (LSTM) network for temporal feature extraction using 2-minute epochs as input; (ii) a CNN to extract spatial features from topological mappings of the independent components; and (iii) a CNN to capture frequency-domain features from power spectral density representations. The experimental setup for evaluating the artifact detection framework is detailed in \Cref{sec:experimental}.

% \newpage
\section{Results and Discussion} \label{sec:results}

Performing a direct validation of the proposed system through a large-scale neonatal trial has considerable practical and ethical challenges at early stages of technology development of such a newly developed device. Therefore, the chosen validation study was designed to avoid neonatal subjects while ensuring that the results we obtained were representative of applications in clinical neurophysiology, leading towards neonatal care. The study was divided into two complementary components:

\begin{enumerate}
    \item \textbf{Hardware performance validation:} Assessment of the EEG acquisition hardware in a representative clinical setting using a pediatric patient.
    \item \textbf{Deep learning validation:} Evaluation of the seizure detection model using a publicly available neonatal cEEG dataset.
\end{enumerate}

The deep learning component establishes the model’s capability to detect seizures from neonatal EEG, while the hardware validation confirms the system’s ability to acquire high-quality EEG signals. Together, these components provide evidence that the integrated system can operate effectively in a clinical setting with few minor improvements.

\subsection{Experimental Setup for Clinical Experiment}
\label{sec:experimental}

EEG signals have low amplitudes in the range of 10--100~\si{\micro\volt} and are approximately 100--1000 times weaker than other physiological biosignals, making them highly susceptible to noise contamination~\cite{teplan2002fundamentals}. Consequently, clinical validation is essential to confirm that the proposed device can acquire high-quality EEG recordings.

\begin{figure}[t!]
    \centering
    \includegraphics[width=1\linewidth]{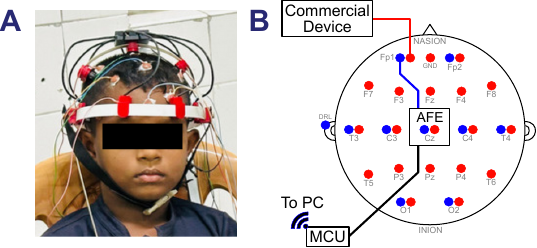}
    \caption{\textbf{Data collection procedure. A) }Image of the participant during the data collection procedure. EEG signals were recorded simultaneously using our custom device and the commercial device available in the hospital. \textbf{B) }Block diagram of setup using both the active dry-contact electrode device and the commercial wet electrode device. Red markers: commercial wet electrodes. Blue markers: active electrodes.}
    \label{fig:clinical}
    \vspace{-10pt}
\end{figure}

\begin{figure*}[tbp]
    \centering
    \includegraphics[width=1\linewidth]{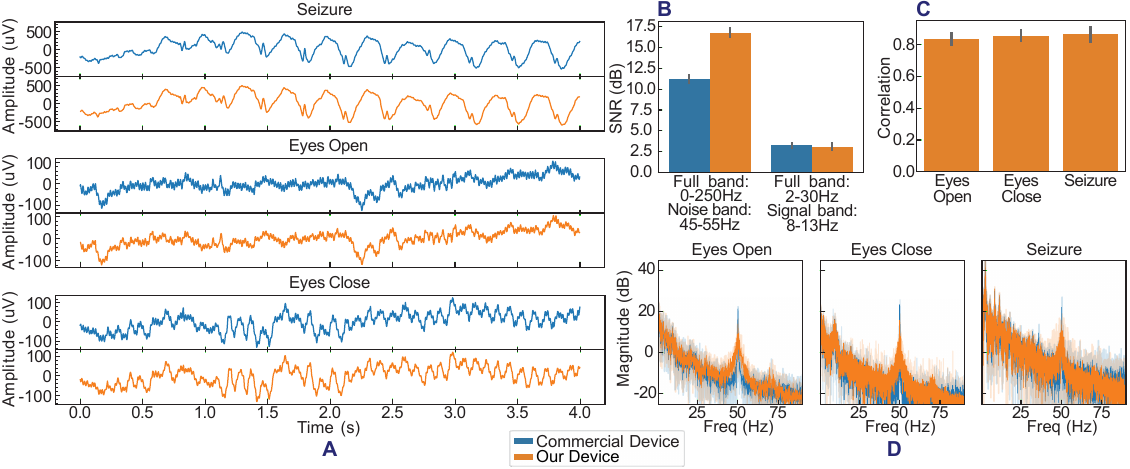}
    \caption{\textbf{Signal quality and signal-to-noise ratio analysis comparing our custom device and the commercial device available in the hospital. A)} Raw unfiltered EEG obtained by both the commercial wet electrode system and the proposed active dry-contact electrode system during various states, showing the high similarity between the signals recorded by the two devices. \textbf{B)} Average SNR obtained by the two systems, calculated separately, considering powerline noise as the noise signal and alpha signal as the true signal. \textbf{C)} Average correlation between the EEG signals obtained by the two systems during various states, showing the high correlation found during each state, with the correlation increasing with the amplitude of the signal. \textbf{D)} Average frequency plots of the two systems during various states, showing the increase in amplitude around 10 Hz during the "Eyes Close" state and a high amplitude in the low frequencies during the "Seizure" state. Further, a slightly higher 50 Hz amplitude (powerline noise) is observed in the commercial device. The gray error bars show the 95\% confidence interval across the dataset.}
    \label{fig:eyes_open}
    \vspace{-10pt}
\end{figure*}

To validate the hardware, our proposed custom dry electrode device was compared against a commercial wet electrode EEG system and tested for functionality during seizure events. For performance benchmarking, visual stimulation using flashing lights at frequencies between 5~Hz and 30~Hz was employed to induce one of the most easily detectable seizure types—absence seizures~\cite{panayiotopoulos1999typical}. Inducing seizures in neonates carries significant risk and potential neurological harm~\cite{kim2023skin}, and performing such experiments on neonatal patients with custom-built hardware poses substantial ethical challenges. To mitigate these risks, validation was conducted using a slightly older pediatric patient volunteer as shown in Fig.~\ref{fig:clinical}A, enabling safe demonstration of the EEG monitoring system’s performance, which has been an approach used in similar studies~\cite{kwak2021skin}.

\begin{table}[t!]
    %\vspace{-11pt}
    \centering
    \caption{Dataset used to calculate signal quality}
    \begin{tabular}{|c|c|c|}
        \hline
         State&  Number of Epochs& Total Length (s)\\
        \hline
         Eyes Opened&  $7$& $41$\\
        \hline
         Eyes Closed&  $6$& $83$\\
        \hline
         Seizure&  $4$&     $58.06$\\
        \hline
    \end{tabular}
    \label{tab:dataset_bc}
    \vspace{-10pt}
\end{table}

Absence seizures are most commonly observed in children between 4 and 10 years of age; therefore, a patient within this age range was selected for the clinical study. The experiment was conducted in collaboration with Lady Ridgeway Hospital for Children (Colombo, Sri Lanka) with ethical approval from the University of Moratuwa, Sri Lanka (Ethics Declaration Number: ERN/2024/001). The electrode montage used in the clinical study followed the configuration described in~\cite{reduced_eeg, ryan2024electrographic}. For direct comparison, EEG was simultaneously recorded using the hospital’s commercial EEG system (Nihon Kohden JE-921, Tokyo, Japan) and our proposed custom device. Electrodes from both systems were positioned in close proximity at corresponding scalp locations to minimize spatial variability in the recordings, as shown in Fig.~\ref{fig:clinical}B.

The patient is an 8-year-old male with a clinical diagnosis of absence seizures. During the recording session, the patient was instructed to perform tasks such as opening and closing the eyes to capture different EEG states. Standard clinical seizure-induction protocols were then followed, including hyperventilation and intermittent photic stimulation~\cite{kane2014hyperventilation}. EEG signals were recorded for approximately 24~minutes, during which four seizure epochs were observed.

% We now present the results of the experiments to verify the effectiveness of the proposed EEG acquisition and seizure detection system. To this end, in \Cref{sec:sqa}, we first analyze the signal quality of our device by comparing it with the commercial device. The results of the seizure detection model and automatic artifact removal algorithm are discussed in \Cref{sec:results_scr} and \Cref{sec:results_arr}, respectively.

\subsection{Signal Quality and Signal-to-Noise Ratio Analysis}
\label{sec:sqa}

\subsubsection{Correlation}

Visual inspection of the raw EEG signals indicated a strong similarity between recordings obtained from the proposed active dry-contact electrode system and the commercial wet-electrode system, as depicted in Fig.~\ref{fig:eyes_open}A. For quantitative assessment, both recordings were bandpass-filtered between 2~Hz and 30~Hz using a Chebyshev type~II filter, and the correlation between the two systems was computed. Details of the samples used for this analysis are provided in \Cref{tab:dataset_bc}.

The average correlation across multiple samples and channels is shown in Fig.~\ref{fig:eyes_open}C. Correlation values were observed to increase during alpha-wave activity and further rise during seizure events. This trend is consistent with the inherent characteristics of EEG signals, where higher-amplitude activity typically yields stronger correlations.

% However, it's important to note that these correlation values surpass those reported in previous studies using dry-electrode-based devices.

\begin{table*}[t!]
\scriptsize
  \centering
  \caption{Model performance comparison. CV - Cross Validation\\\textcolor{orange}{Orange} highlights the proposed model, while \textcolor{blue}{blue} indicates previous studies on ML for reduced montages, which lack generalizability to unseen subjects as they are trained and tested in a subject-specific configuration.}
  \label{tab:Result_table}
  \renewcommand{\arraystretch}{1.5}
  \begin{tabular}{|>{\centering\arraybackslash}p{1.7cm}|>{\centering\arraybackslash}p{2.35cm}|>{\centering\arraybackslash}p{1.1cm}|c|c|c|c|c|c|}
    \hline
    Number of EEG & \multirow{2}{*}{\begin{tabular}{c}
          {}Method
          \end{tabular}}& \multirow{2}{*}{\parbox[c]{1.1cm}{\centering Dataset}} & Accuracy& AUC& \multirow{2}{*}{\parbox[c]{1.1cm}{\centering Specificity\\/FPR$^*$}}& \multirow{2}{*}{\begin{tabular}{c}
          {}Recall
     \end{tabular}} & \multirow{2}{*}{\begin{tabular}{c}
          {}Precision
     \end{tabular}} & \multirow{2}{*}{\begin{tabular}{c}
          {}Kappa
     \end{tabular}}\\ \cline{4-5}
     channels& & & mean$\pm$ std & Median (IQR)&& & &\\ \hline

     \multirow{7}{*}{\begin{tabular}{c}
          {}$\geq18^{1}$
     \end{tabular}}
     &ST-GAT (FL) \cite{raeisi2023class}& \multirow{4}{*}{\parbox[c]{1.1cm}{\centering Helsinki\\\cite{stevenson2019dataset}\\(Neonatal)}}&-&$99.30$ $(96.40,99.50)$&$0.86^*$&$\mathbf{98.00}$&-&$0.88$\\ 
     &XL ConvNeXt \cite{hogan2025scaling}&&-&$99.6 (97.5, 1.00)$&$0.34^*$&-&-&$0.80$\\ 
     &LMA-EEGNet \cite{zhou2024lma}&&$95.7$&$98.62$&$96.4$&$95.0$&-&-\\
     & Wang \textit{et al.} \cite{wang2024combining}&&$92.34$&-&-&$98.74$&$93.61$&-\\ \cline{2-9}
     & DCAE+Bi-LSTM\cite{abdelhameed2021deep} &\multirow{3}{*}{\parbox[c]{1.1cm}{\centering CHB-MIT\\\cite{chbmit}\\(1.5 - 22 years)}}& $98.79\pm0.53$&-&$98.86$&$98.72$&$98.86$&-\\
     & Zhang \textit{et al.} \cite{zhang2024scheme}&&$99.35$&$99.34$&$99.51$&$99.24$&-&-\\
     & Li \textit{et al.} \cite{li2025cnn}&& $98.54$ &-&$98.55$&$99.54$&-&-\\     \hline

   $12$ &ST-GAT (FL)$^2$&\multirow{4}{*}{\parbox[c]{1.1cm}{\centering Helsinki\\\cite{stevenson2019dataset}}}&$80.29$$\pm$$9.48$&$83.98$ $(77.80,90.90)$ &-&$39.98$&$\mathbf{94.91}$&$0.43$\\ 
    $10$-fold CV&\textcolor{orange}{Our method}&&\textcolor{orange}{$\mathbf{89.02}\pm2.91$}&\textcolor{orange}{$\mathbf{91.84}$ $ (88.57,95.21)$} &-&\textcolor{orange}{$\mathbf{82.84}$}&\textcolor{orange}{$94.23$}&\textcolor{orange}{$\mathbf{0.89}$}\\ \cline{1-2}\cline{4-9}
    $12$ &ST-GAT (FL)$^2$&&$88.80$&$91.71$&-&$66.89$&$\mathbf{95.17}$&$0.71$\\
    $(80\%-20\%)$&\textcolor{orange}{Our method}&&\textcolor{orange}{$\mathbf{91.56}$}&\textcolor{orange}{$\mathbf{94.42}$}&\textcolor{orange}{$90.86$}&\textcolor{orange}{$\mathbf{83.22}$}&\textcolor{orange}{$88.61$}&\textcolor{orange}{$\mathbf{0.80}$}\\ \hline
    $10$&\textcolor{blue}{Asif \textit{et al.} \cite{reduced_mon}}&\multirow{2}{*}{\parbox[c]{1.1cm}{\centering CHB-MIT\\\cite{chbmit}}}&\textcolor{blue}{$99.62$}&-&\textcolor{blue}{$99.7$}&\textcolor{blue}{$92.4$}&-&-\\ \cline{1-2}\cline{4-9}
    $12$&\textcolor{blue}{Selvakumari \textit{et al.} \cite{selvakumari2019patient}}&&\textcolor{blue}{$95.63$}&-&\textcolor{blue}{$96.55$}&\textcolor{blue}{$95.7$}&-&-\\ \hline
    \multicolumn{9}{l}{}\\
    \multicolumn{9}{l}{$^{1}$These models are trained with full montage EEG data with 18 or more EEG channels.  }\\
    \multicolumn{9}{l}{$^2$We retrained ST-GAT (FL), the best variant of the ST-GAT and the counterpart of the proposed model, for the 12-channel reduced montage and report the evaluated}\\
    \multicolumn{9}{l}{results  to show a fair comparison between our model and the current best SOTA model.}
  \end{tabular}
  \vspace{-5pt}
\end{table*}

\subsubsection{Signal-to-Noise Ratio}

To assess signal quality, the signal-to-noise ratio (SNR) was calculated following an analysis of the frequency spectrum of the acquired signals. The frequency spectra of the raw EEG recordings obtained under various physiological states are compared in Fig.~\ref{fig:eyes_open}D. A slightly higher 50~Hz powerline component is evident in the recordings from the commercial wet-electrode system.

Estimating SNR in EEG is challenging due to the difficulty of isolating pure noise. Previous studies have approximated SNR by comparing the power of the alpha band (8--13~Hz) to the power of the 2--30~Hz bandpass-filtered EEG~\cite{chen2014soft}. However, this method is limited because EEG activity extends beyond the 2--30~Hz range. To address this, two approaches were evaluated:

\begin{enumerate}
    \item \textbf{Powerline frequency as noise:} The 50~Hz powerline component was treated as the primary noise source within the 2--90~Hz bandpass-filtered signal.
    \item \textbf{Alpha wave as signal:} The alpha band was considered the signal of interest in the 2--30~Hz bandpass-filtered signal, with all other components regarded as noise.
\end{enumerate}

The SNR results for both methods are shown in Fig.~\ref{fig:eyes_open}B. Across all estimation approaches, the proposed device achieved SNR values that were consistently higher or comparable to those of the commercial wet-electrode system.
\vspace{5pt}

\subsubsection{Noise During Movements}

Dry-contact electrode EEG systems are known to be susceptible to motion artifacts~\cite{searle2000direct}. In the proposed device, minimal signal variation was observed during small movements, such as breathing; however, noise disruptions occurred when the subject fell asleep and exhibited larger head movements.

Notably, the O1 and O2 electrodes positioned on the forehead demonstrated motion-noise levels comparable to those of the wet-electrode system, even during larger movements, indicating that the core acquisition technology is robust. In contrast, electrodes with looser scalp contact exhibited brief signal disruptions during motion, likely due to the intentional reduction in contact pressure to enhance user comfort. Future design iterations may address this trade-off by incorporating spring-loaded mechanisms or flexible conductive materials to improve electrode stability while maintaining comfort.

\subsection{Seizure Classification Results} 
\label{sec:results_scr}

\begin{figure*}[t!]
    \centering
    \includegraphics[width=0.95\linewidth]{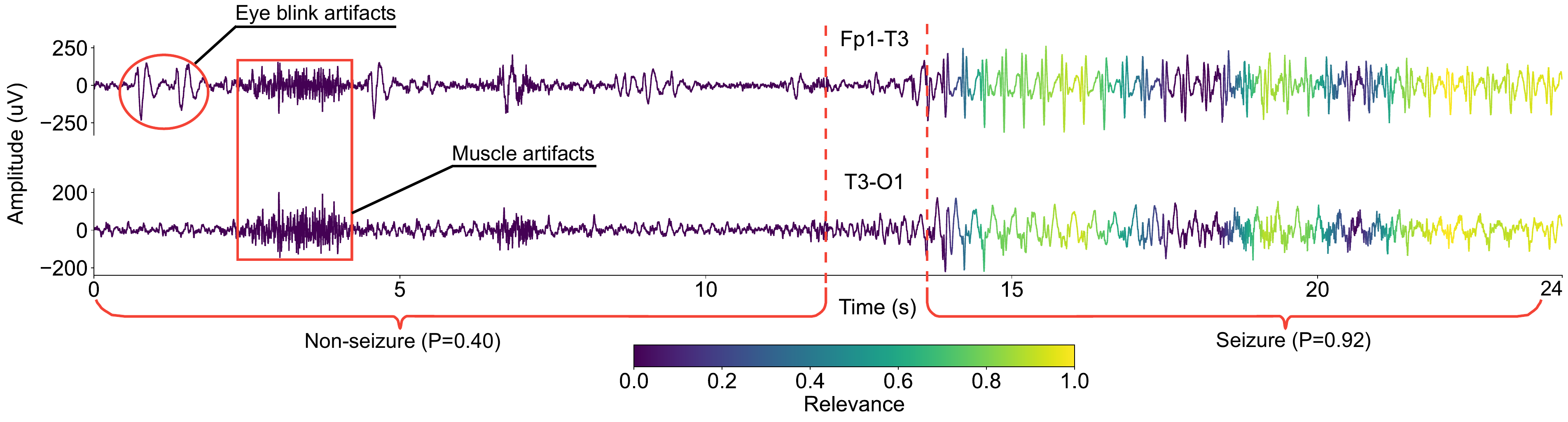}
    \caption{\textbf{Deep learning inference results on data collected from the clinical experiment at Lady Ridgeway Hospital, Sri Lanka}. Fp1-T3 and T3-O1 channels are visualized here, depicting an EEG recording with the first 12 s representing a normal EEG epoch with a probability of 0.4 to be a seizure and the last 12 s representing a seizure epoch with a probability of 0.92 to be a seizure. The onset of a seizure can be seen between the two red dash lines (12--13.5 s). Although artifacts are similar in amplitude to a seizure, they don't impact the detection probability.}
    \label{fig:ML_results}
    % \vspace{-10pt}
\end{figure*}

\subsubsection{Dataset Preparation for Training and Evaluation}
The publicly available Helsinki neonatal seizure EEG dataset~\cite{stevenson2019dataset} was used to train and evaluate the proposed machine learning model. Each EEG recording was segmented into 12-second epochs after extracting the signals for each bipolar channel, as indicated by the blue arrows in Fig.~\ref{fig:headset_design}A. The data preparation followed the procedure described in \cite{AI_paper}, employing overlapping segments of 11 seconds for seizure epochs and 10 seconds for non-seizure epochs to reduce class imbalance. An epoch was labeled as seizure (positive) if it contained at least 1 second of seizure activity; otherwise, it was labeled as non-seizure (negative). 

To the best of our knowledge, this is the \emph{first study} to investigate an explainable deep learning model for neonatal seizure detection using a 12-channel reduced montage. While \Cref{tab:Result_table} compares the proposed approach with several previous state-of-the-art (SOTA) methods~\cite{raeisi2023class,hogan2025scaling,zhou2024lma,wang2024combining,abdelhameed2021deep,zhang2024scheme,li2025cnn} that utilize 18 or more channels, these prior results were obtained using different datasets and montage configurations. According to the literature, the Helsinki dataset~\cite{stevenson2019dataset} is the only publicly available neonatal EEG dataset. In contrast, the CHB-MIT dataset~\cite{chbmit} contains data from older subjects, with ages ranging from 1.5 to 22~years.

\subsubsection{Model Evaluation and Comparison with State-of-the-Art}
To ensure a fair performance comparison, the counterpart model ST-GAT (FL)~\cite{raeisi2023class} was retrained for 12-channel input using the same dataset and evaluation strategy as the proposed method. As shown in \Cref{tab:Result_table}, the proposed model outperformed the retrained ST-GAT (FL) across multiple metrics, including accuracy, area under the curve (AUC), recall, and Cohen’s kappa. In particular, the proposed model achieved the highest kappa value of 0.89 in 10-fold cross-validation, irrespective of the number of channels.
Previous studies have explored seizure detection using reduced montages, such as the RUSBoost-based method in \cite{reduced_mon} and the support vector machine (SVM)-based method in~\cite{selvakumari2019patient}. Although these models reported improved results, their evaluation was conducted under subject-specific configurations, limiting their generalizability to unseen patients. Similarly, other recent studies~\cite{abdelhameed2021deep, zhang2024scheme, li2025cnn} also employed subject-specific training, which does not reflect the more challenging scenario of cross-subject generalization addressed in this work.

\subsubsection{Real-Time Deployment and Interpretability}
The proposed deep learning model was integrated into a GUI to enable real-time seizure detection. The system was evaluated qualitatively on real-time EEG data collected during the clinical experiment. The model requires observing at least 5 seconds of seizure activity within a 12-second epoch to classify it as a true positive. As illustrated in Fig.~\ref{fig:ML_results}, the model accurately classified EEG epochs contaminated with high-amplitude artifacts, such as muscle activity and eye movements, as non-seizure. Seizure patterns exhibiting high amplitude were also correctly identified. However, in some of the seizure events, the qualitative analysis did reveal occasional false positives caused by artifacts with very high amplitudes. The number of false negatives was negligible. 

To enhance transparency in decision-making, a real-time interpretability framework~\cite{AI_paper} based on a modified Grad-CAM~\cite{grad-cam} was incorporated. This approach highlights temporal regions and channels most relevant to the model’s output, providing clinicians with insights into the model’s reasoning process. As shown in Fig.~\ref{fig:ML_results}, during seizure epochs, the model assigns higher relevance scores to specific regions of the EEG, thereby improving user trust and facilitating adoption in practical settings.

\subsection{Artifact Removal Results}
\label{sec:results_arr}

A modified implementation of the IClabel model architecture in~\cite{lopes2022ensemble}, incorporating an additional batch normalization layer, was trained on the EPIC dataset~\cite{lopes-2022}. The performance of this modified model ($\text{IClabel}_{\mathrm{our}}$) was compared with the original implementation ($\text{IClabel}_{\mathrm{en}}$) in Table~\ref{tab:Result_table_IC}, based on overall accuracy, area under the curve (AUC), F1-score, precision, and recall across different classes. The proposed model achieved an overall accuracy of 91.96\% and an AUC of 91.85\%, representing a modest improvement over the baseline. Class-wise evaluation revealed a high F1-score of 92\% for non-artifactual instances, demonstrating the model’s capability to accurately distinguish clean EEG signals. For the artifactual class, the model achieved a precision of 88\% and a recall of 91\%. In comparison, the $\text{IClabel}_{\mathrm{en}}$ model obtained an overall accuracy of 90.09\% and an AUC of 90.90\%, with similar trends observed in the per-class metrics.

\begin{figure}[t!]
    \centering
    % \includegraphics[width=1\linewidth]{figures/Artefact combined.pdf}
    % \\[10pt]
    \includegraphics[width=1\linewidth]{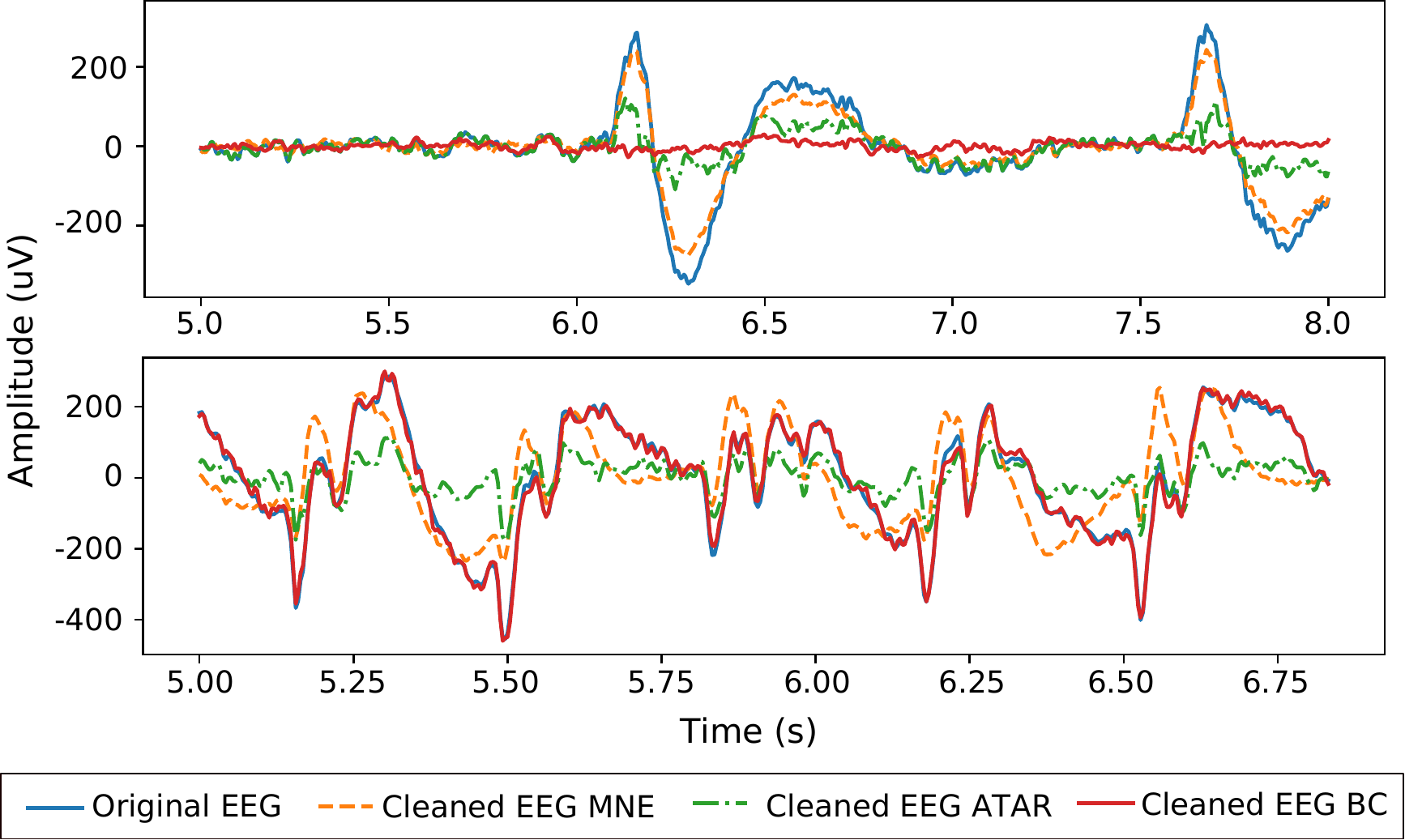} 
    \caption{\textbf{Comparison of different artifact removal algorithms with our method.} Top - An instance of an eye-blink artifact. Bottom - An instance of a seizure. Our method effectively removes artifacts while retaining seizure activity, outperforming other approaches.}
    \label{fig:nirmana2}
    \vspace{-15pt}
\end{figure}

The modified algorithm was further evaluated on data acquired during the clinical experiment at Lady Ridgeway Hospital, with results presented in Fig.~\ref{fig:nirmana2}. The inclusion of the batch normalization layer enhanced the model’s ability to remove EEG artifacts while preserving seizure-related patterns, outperforming both the MNE-ICALabel~\cite{Li2022} and ATAR~\cite{bajaj2020automatic} algorithms. MNE-ICALabel failed to detect the artifact, likely due to limitations in its training set, whereas ATAR’s wavelet-based filtering approach removed both the artifact and the seizure pattern, reflecting its limited context sensitivity. In contrast, the proposed model successfully differentiated artifacts from seizure activity by leveraging spatial, temporal, and frequency-domain features. % ELABORATE A LITTLE MORE

\setlength{\tabcolsep}{2pt}

\begin{table}[t!]
% \scriptsize
  \centering
  \caption{Model performance ($\%$) comparison of the IC label model}
  \label{tab:Result_table_IC}
  \begin{tabular}{ccccccc}
\hline
\multicolumn{1}{c}{{Model}} & \multicolumn{1}{c}{{Class}} & \multicolumn{1}{c}{{Accuracy}} & \multicolumn{1}{c}{{AUC}} & \multicolumn{1}{c}{{F1}} & \multicolumn{1}{c}{{Precision}} & \multicolumn{1}{c}{{Recall}} \\ \hline
\multirow{3}{*}{\begin{tabular}[c]{@{}l@{}}$\text{IClabel}_{our}$\end{tabular}} &  & \multicolumn{1}{c}{$91.96$} & \multicolumn{1}{c}{$91.85$} & $92$ & \multicolumn{1}{c}{-} & \multicolumn{1}{c}{-} \\ 
 & Artifactual & - & - & $89$& $88$ & $91$ \\ 
 & Non-Artifactual & - & - & $93$ & $95$ & $92$ \\ \hline
\multirow{3}{*}{\begin{tabular}[c]{@{}l@{}}$\text{IClabel}_{en}$ \end{tabular}} &  & \multicolumn{1}{c}{$90.09$} & \multicolumn{1}{c}{$90.90$} & $90$ & \multicolumn{1}{c}{-} & \multicolumn{1}{c}{-} \\
 & Artifactual & - & - & $88$& $82$ & $94$ \\  
 & Non-Artifactual & - & - & $92$& $96$& $88$ \\ \hline
\end{tabular}
\vspace{-10pt}
\end{table}

%\textcolor{blue}{\textbf{NIRMANA check above stuff. add about results from hospital and write about MNE ATAR stuff}}%

\section{Conclusion and Future Work} 
\label{sec:conc}

In this study, a novel, low-cost, active dry-contact electrode-based EEG system was presented for the detection of seizures in a representative clinical setting. The proposed approach addresses key limitations of conventional wet-electrode systems, including lengthy setup times and the discomfort associated with conductive gels. By employing a clinically validated reduced-montage EEG headset, the system demonstrated performance comparable to a gold-standard wet-electrode device, while improving usability for both caregivers and patients at a fraction of the cost of a commercial cEEG system. The proposed seizure detection model achieved a 2.76\% improvement in accuracy and a 16.33\% improvement in recall over state-of-the-art methods. In addition, an integrated artifact removal algorithm was shown to effectively suppress artifacts while preserving seizure-related patterns. These findings highlight the potential of the proposed EEG system to enable continuous monitoring in both clinical and home environments, offering a practical and accessible tool for enhancing care in clinical neurophysiology. 

Although the hardware performance evaluation in this study was limited to pediatric patients above the neonatal age group, future work will focus on adapting the system specifically for neonatal use. This will include the development of a customized dry-electrode neonatal headcap with adjustable sizing, as well as soft-contact electrodes without spikes, designed to ensure biocompatibility and minimize the risk of skin injury for neonates. Once these refinements are complete and renewed ethical approval is obtained, a subsequent hardware performance study will be conducted directly on neonatal patients to validate the effectiveness of this system for EEG monitoring within NICUs. 

In practice, a significant challenge in the use of dry-contact electrodes is the occurrence of motion artifacts, which hinder the acquisition of reliable EEG recordings unless the subject remains perfectly still; an impractical requirement, particularly for infants, babies, and young children. Such movements cause variations in the electrode–skin impedance encountered by the dry electrodes, which can degrade signal quality. Although the present work does not implement a real-time strategy for motion artifact cancellation, we are investigating an integrated solution combining circuit-level enhancements and signal processing techniques to enable real-time motion artifact cancellation. Future work, such as that described above, will focus on further system optimization and broader clinical validation, with the goal of developing a robust solution for low-cost, active dry-contact electrode-based EEG monitoring within NICUs.

\section*{Acknowledgement}

The computational resources used in the project were funded by the Accelerating Higher Education Expansion and Development (AHEAD) Operation of the Ministry of Higher Education of Sri Lanka, funded by the World Bank. We thank the staff and technicians from Lady Ridgeway Hospital for Children, Sri Lanka, for facilitating the patient experiments.

\setlength{\biblabelsep}{0.5em}
\AtNextBibliography{\footnotesize\selectfont}

\printbibliography

@article{lee2018dry,
  title={Dry electrode-based fully isolated {EEG/fNIRS} hybrid brain-monitoring system},
  author={Lee, Seungchan and Shin, Younghak and Kumar, Anil and Kim, Minhee and Lee, Heung-No},
  journal={IEEE Trans. Biomed. Eng.},
  volume={66},
  number={4},
  pages={1055--1068},
  year={2018},
  publisher={IEEE}
}

@inproceedings{sullivan2008brain,
  title={A brain-machine interface using dry-contact, low-noise {EEG} sensors},
  author={Sullivan, Thomas J and Deiss, Stephen R and Jung, Tzyy-Ping and Cauwenberghs, Gert},
  booktitle={2008 IEEE International Symposium on Circuits and Systems (ISCAS)},
  pages={1986--1989},
  year={2008},
  organization={IEEE}
}

@article{huang2014novel,
  title={Novel active comb-shaped dry electrode for {EEG} measurement in hairy site},
  author={Huang, Yan-Jun and Wu, Chung-Yu and Wong, Alice May-Kuen and Lin, Bor-Shyh},
  journal={IEEE Trans. Biomed. Eng.},
  volume={62},
  number={1},
  pages={256--263},
  year={2014},
  publisher={IEEE}
}

@article{pourahmad2016evaluation,
  title={Evaluation of a low-cost and low-noise active dry electrode for long-term biopotential recording},
  author={Pourahmad, Ali and Mahnam, Amin},
  journal={Journal of Medical Signals \& Sensors},
  volume={6},
  number={4},
  pages={197--202},
  year={2016},
  publisher={Medknow}
}

@article{chuang2019cost,
  title={Cost-efficient, portable, and custom multi-subject electroencephalogram recording system},
  author={Chuang, Kai-Chiang and Lin, Yuan-Pin},
  journal={IEEE Access},
  volume={7},
  pages={56760--56769},
  year={2019},
  publisher={IEEE}
}

@article{kubota2018continuous,
  title={Continuous {EEG} monitoring in {ICU}},
  author={Kubota, Yuichi and Nakamoto, Hidetoshi and Egawa, Satoshi and Kawamata, Takakazu},
  journal={Journal of Intensive Care},
  volume={6},
  pages={1--8},
  year={2018},
  publisher={Springer}
}

@article{fonseca2006novel,
  title={A novel dry active electrode for {EEG} recording},
  author={Fonseca, Carlos and Cunha, JP Silva and Martins, Rui Escadas and Ferreira, Victor M and De Sa, JP Marques and Barbosa, MA and Da Silva, A Martins},
  journal={{IEEE Trans. Biomed. Eng.}},
  volume={54},
  number={1},
  pages={162--165},
  year={2006},
  publisher={IEEE}
}

@article{lopes2022ensemble,
  title={Ensemble deep neural network for automatic classification of {EEG} independent components},
  author={Lopes, F{\'a}bio and Leal, Adriana and Medeiros, J{\'u}lio and Pinto, Mauro F and Dourado, Antonio and D{\"u}mpelmann, Matthias and Teixeira, C{\'e}sar},
  journal={IEEE Trans. Neural Syst. Rehabil. Eng},
  volume={30},
  pages={559--568},
  year={2022},
  publisher={IEEE}
}

@article{shad2020impedance,
  title={Impedance and noise of passive and active dry {EEG} electrodes: a review},
  author={Shad, Erwin Habibzadeh Tonekabony and Molinas, Marta and Ytterdal, Trond},
  journal={IEEE Sensors J.},
  volume={20},
  number={24},
  pages={14565--14577},
  year={2020},
  publisher={IEEE}
}

@article{panayiotopoulos1999typical,
  title={Typical absence seizures and their treatment},
  author={Panayiotopoulos, CP},
  journal={Archives of disease in childhood},
  volume={81},
  number={4},
  pages={351--355},
  year={1999},
  publisher={BMJ Publishing Group Ltd}
}

@article{kane2014hyperventilation,
  title={Hyperventilation during electroencephalography: safety and efficacy},
  author={Kane, Nick and Grocott, Lesley and Kandler, Ros and Lawrence, Sarah and Pang, Catherine},
  journal={Seizure},
  volume={23},
  number={2},
  pages={129--134},
  year={2014},
  publisher={Elsevier}
}

@INPROCEEDINGS{AI_paper,
  author={Udayantha, Dinuka Sandun and Weerasinghe, Kavindu and Wickramasinghe, Nima and Abeyratne, Akila and Wickremasinghe, Kithmin and Wanigasinghe, Jithangi and De Silva, Anjula and Edussooriya, Chamira U. S.},
  booktitle={2024 IEEE International Conference on Systems, Man, and Cybernetics (SMC)}, 
  title={Using Explainable {AI} for {EEG}-based Reduced Montage Neonatal Seizure Detection}, 
  year={2024},
  volume={},
  number={},
  pages={463-468},
 }

@article{electrode_material,
    author = "Popović-Maneski, L. and Ivanović, M.D. and Atanasoski, V. and Miletić, M. and Zdolšek, S. and Bojović, B. and Hadžievski, L.",
    title = "Properties of different types of dry electrodes for wearable smart monitoring devices",
    journal = "Biomed Tech (Berl)",
    volume = "65",
    number = "4",
    pages = "405-415",
    year = "2020",
    DOI = "10.1515/bmt-2019-0167",
}

@article{electrode_shape,
    author = "Di Flumeri, G. and Aricò, P. and Borghini, G. and Sciaraffa, N. and Di Florio, A. and Babiloni, Fabio ",
    title = "The Dry Revolution: Evaluation of Three Different {EEG} Dry Electrode Types in Terms of Signal Spectral Features, Mental States Classification and Usability",
    journal = "Sensors",
    volume = "19",
    number = "6",
    pages = "1365",
    year = "2019",
    DOI = "https://doi.org/10.3390/s19061365",
}

@article{active_shield,
author = {Jiang, Yanbing and Samuel, Oluwarotimi and Liu, Xueyu and Wang, Xin and Idowu, Oluwagbenga and Li, Peng and Chen, Fei and Zhu, Mingxing and Geng, Yanjuan and Chen, Shixiong and Li, Patrick},
year = {2018},
month = {02},
pages = {},
title = {Effective Biopotential Signal Acquisition: Comparison of Different Shielded Drive Technologies},
volume = {8},
number = {2},
journal = {Applied Sciences},
doi = {10.3390/app8020276}
}

@inproceedings{resnet,
  title={Deep residual learning for image recognition},
  author={He, Kaiming and Zhang, Xiangyu and Ren, Shaoqing and Sun, Jian},
  booktitle={Proceedings of the IEEE conference on computer vision and pattern recognition},
  pages={770--778},
  year={2016}
}

@INPROCEEDINGS{grad-cam,
  author={Selvaraju, Ramprasaath R. and Cogswell, Michael and Das, Abhishek and Vedantam, Ramakrishna and Parikh, Devi and Batra, Dhruv},
  booktitle={2017 IEEE International Conference on Computer Vision (ICCV)}, 
  title={{Grad-CAM}: Visual Explanations from Deep Networks via Gradient-Based Localization}, 
  year={2017},
  volume={},
  number={},
  pages={618-626},
  doi={10.1109/ICCV.2017.74}}

@article{stevenson2019dataset,
  title={A dataset of neonatal EEG recordings with seizure annotations},
  author={Stevenson, Nathan J and Tapani, Karoliina and Lauronen, Leena and Vanhatalo, Sampsa},
  journal={Scientific data},
  volume={6},
  number={1},
  pages={1--8},
  year={2019},
  publisher={Nature Publishing Group}
}

@article{raeisi2023class,
  title={A class-imbalance aware and explainable spatio-temporal graph attention network for neonatal seizure detection},
  author={Raeisi, Khadijeh and Khazaei, Mohammad and Tamburro, Gabriella and Croce, Pierpaolo and Comani, Silvia and Zappasodi, Filippo and others},
  journal={International journal of neural systems},
  volume={33},
  number={9},
  pages={2350046},
  year={2023}
}

@ARTICLE{reduced_mon,
  author={Asif, Raheel and Saleem, Sajid and Hassan, Syed Ali and Alharbi, Soltan Abed and Kamboh, Awais Mehmood},
  journal={IEEE Access}, 
  title={Epileptic Seizure Detection With a Reduced Montage: A Way Forward for Ambulatory {EEG} Devices}, 
  year={2020},
  volume={8},
  number={},
  pages={65880-65890},
  doi={10.1109/ACCESS.2020.2983917}}

@article{selvakumari2019patient,
  title={Patient-specific seizure detection method using hybrid classifier with optimized electrodes},
  author={Selvakumari, R Shantha and Mahalakshmi, M and Prashalee, P},
  journal={Journal of medical systems},
  volume={43},
  number={5},
  pages={121--127},
  year={2019},
  publisher={Springer}
}

@article{teplan2002fundamentals,
  title={Fundamentals of {EEG} measurement},
  author={Teplan, Michal and others},
  journal={Measurement science review},
  volume={2},
  number={2},
  pages={1--11},
  year={2002}
}

@article{lopez2014dry,
  title={Dry {EEG} electrodes},
  author={Lopez-Gordo, Miguel Angel and Sanchez-Morillo, Daniel and Valle, F Pelayo},
  journal={Sensors},
  volume={14},
  number={7},
  pages={12847--12870},
  year={2014},
  publisher={MDPI}
}

@article{mizrahi1987characterization,
  title={Characterization and classification of neonatal seizures},
  author={Mizrahi, Eli M and Kellaway, Peter},
  journal={Neurology},
  volume={37},
  number={12},
  pages={1837--1837},
  year={1987},
  publisher={AAN Enterprises}
}

@incollection{krawiec2023neonatal,
  title={Neonatal seizure},
  author={Krawiec, Conrad and Muzio, Maria Rosaria},
  booktitle={StatPearls [Internet]},
  year={2023},
  publisher={StatPearls Publishing}
}

@article{reduced_eeg,
author = {Tekgul, Hasan and Bourgeois, Blaise F.D. and Gauvreau, Kimberlee and Bergin,Ann M },
year = {2005},
month = {03},
pages = {155-161},
title = {Electroencephalography in neonatal seizures: comparison of a reduced and a full 10/20 montage},
volume = {32},
number = {3},
journal = {Pediatric Neurology},
doi = {10.1016/j.pediatrneurol.2004.09.014}
}

@article{searle2000direct,
  title={A direct comparison of wet, dry and insulating bioelectric recording electrodes},
  author={Searle, Andrew and Kirkup, LJPM},
  journal={Physiological measurement},
  volume={21},
  number={2},
  pages={271--283},
  year={2000},
  publisher={IOP Publishing}
}

@article{murray2008defining,
  title={Defining the gap between electrographic seizure burden, clinical expression and staff recognition of neonatal seizures},
  author={Murray, Deirdre M and Boylan, Geraldine B and Ali, Imhad and Ryan, C Anthony and Murphy, Brendan P and Connolly, Sean},
  journal={Archives of Disease in Childhood-Fetal and Neonatal Edition},
  volume={93},
  number={3},
  pages={F187--F191},
  year={2008},
  publisher={BMJ Publishing Group}
}

@article{Li2022,
  title = {{MNE-ICALabel}: Automatically annotating {ICA} components with {ICLabel} in Python},
  volume = {7},
  ISSN = {2475-9066},
  url = {http://dx.doi.org/10.21105/joss.04484},
  DOI = {10.21105/joss.04484},
  number = {76},
  journal = {Journal of Open Source Software},
  publisher = {The Open Journal},
  author = {Li,  Adam and Feitelberg,  Jacob and Saini,  Anand Prakash and H\"{o}chenberger, Richard and Scheltienne,  Mathieu},
  year = {2022},
  pages = {4484--4486},
  month = aug,
  pages = {4484}
}

@article{chen2014soft,
  title={Soft, comfortable polymer dry electrodes for high quality {ECG} and {EEG} recording},
  author={Chen, Yun-Hsuan and De Beeck, Maaike Op and Vanderheyden, Luc and Carrette, Evelien and Mihajlovi{\'c}, Vojkan and Vanstreels, Kris and Grundlehner, Bernard and Gadeyne, Stefanie and Boon, Paul and Van Hoof, Chris},
  journal={Sensors},
  volume={14},
  number={12},
  pages={23758--23780},
  year={2014},
  publisher={MDPI}
}

@article{lee1999independent,
  title={Independent component analysis using an extended infomax algorithm for mixed subgaussian and supergaussian sources},
  author={Lee, Te-Won and Girolami, Mark and Sejnowski, Terrence J},
  journal={Neural computation},
  volume={11},
  pages={417--441},
  year={1999},
  publisher={MIT Press}
}

@article{lopes-2022,
	author = {Lopes, Fábio and Leal, Adriana and Medeiros, Júlio and Pinto, Mauro F. and Dourado, António and Dümpelmann, Matthias and Teixeira, César},
	journal = {Scientific data},
	month = {8},
	number = {1},
	title = {{EPIC: Annotated epileptic {EEG} independent components for artifact reduction}},
	volume = {9},
    pages ={512},
	year = {2022},
	doi = {10.1038/s41597-022-01524-x},
	url = {https://www.nature.com/articles/s41597-022-01524-x},
}

@article{bajaj2020automatic,
  title={Automatic and tunable algorithm for {EEG} artifact removal using wavelet decomposition with applications in predictive modeling during auditory tasks},
  author={Bajaj, Nikesh and Carri{\'o}n, Jes{\'u}s Requena and Bellotti, Francesco and Berta, Riccardo and De Gloria, Alesandro},
  journal={Biomedical Signal Processing and Control},
  volume={55},
  pages={101624},
  year={2020},
  publisher={Elsevier}
}

@article{hauser1992seizure,
  title={Seizure disorders: the changes with age},
  author={Hauser, W Allen},
  journal={Epilepsia},
  volume={33},
  pages={6--14},
  year={1992},
  publisher={Wiley Online Library}
}

@article{wanigasinghe2024seizures,
  title={Seizures in small brains},
  author={Wanigasinghe, Jithangi},
  journal={Sri Lanka Journal of Neurology},
  volume={11},
  pages={3--8},
  number={1},
  year={2024}
}

@article{temko2011eeg,
  title={{EEG}-based neonatal seizure detection with support vector machines},
  author={Temko, Andriy and Thomas, Eoin and Marnane, William and Lightbody, Gordon and Boylan, G},
  journal={Clinical Neurophysiology},
  volume={122},
  number={3},
  pages={464--473},
  year={2011},
  publisher={Elsevier}
}

@article{hossain2019applying,
  title={Applying deep learning for epilepsy seizure detection and brain mapping visualization},
  author={Hossain, M Shamim and Amin, Syed Umar and Alsulaiman, Mansour and Muhammad, Ghulam},
  journal={ACM Transactions on Multimedia Computing, Communications, and Applications (TOMM)},
  volume={15},
  number={1s},
  pages={1--17},
  year={2019},
  publisher={ACM New York, NY, USA}
}

@article{chi2010dry,
  title={Dry-contact and noncontact biopotential electrodes: Methodological review},
  author={Chi, Yu Mike and Jung, Tzyy-Ping and Cauwenberghs, Gert},
  journal={IEEE Rev. Biomed. Eng.},
  volume={3},
  pages={106--119},
  year={2010},
  publisher={IEEE}
}

@article{hogan2025scaling,
  title={Scaling convolutional neural networks achieves expert level seizure detection in neonatal {EEG}},
  author={Hogan, Robert and Mathieson, Sean R and Luca, Aurel and Ventura, Soraia and Griffin, Sean and Boylan, Geraldine B and O’Toole, John M},
  journal={npj Digital Medicine},
  volume={8},
  number={1},
  pages={17--27},
  year={2025},
  publisher={Nature Publishing Group UK London}
}

@ARTICLE{10675445,
  author={Abdallah, Tala and Jrad, Nisrine and Hajjar, Sally El and Abdallah, Fahed and Humeau-Heurtier, Anne and Howayek, Eliane El and Van Bogaert, Patrick},
  journal={IEEE Trans. Biomed. Eng.}, 
  title={Deep Clustering for Epileptic Seizure Detection}, 
  year={2025},
  volume={72},
  number={2},
  pages={480-492},
  doi={10.1109/TBME.2024.3458177}}

@article{ryan2024electrographic,
  title={Electrographic monitoring for seizure detection in the neonatal unit: current status and future direction},
  author={Ryan, Mary Anne J and Malhotra, Atul},
  journal={Pediatric research},
  volume={96},
  number={4},
  pages={896--904},
  year={2024},
  publisher={Nature Publishing Group US New York}
}

@article{abdelhameed2021deep,
  title={A deep learning approach for automatic seizure detection in children with epilepsy},
  author={Abdelhameed, Ahmed and Bayoumi, Magdy},
  journal={Frontiers in Computational Neuroscience},
  volume={15},
  pages={650050},
  year={2021},
  publisher={Frontiers Media SA}
}

@article{chbmit,
  title={PhysioBank, PhysioToolkit, and PhysioNet: components of a new research resource for complex physiologic signals},
  author={Goldberger, Ary L and Amaral, Luis AN and Glass, Leon and Hausdorff, Jeffrey M and Ivanov, Plamen Ch and Mark, Roger G and Mietus, Joseph E and Moody, George B and Peng, Chung-Kang and Stanley, H Eugene},
  journal={circulation},
  volume={101},
  number={23},
  pages={e215--e220},
  year={2000},
  publisher={Lippincott Williams \& Wilkins}
}

@article{zhang2024scheme,
  title={A scheme combining feature fusion and hybrid deep learning models for epileptic seizure detection and prediction},
  author={Zhang, Jincan and Zheng, Shaojie and Chen, Wenna and Du, Ganqin and Fu, Qizhi and Jiang, Hongwei},
  journal={Scientific Reports},
  volume={14},
  number={1},
  pages={16916--16931},
  year={2024},
  publisher={Nature Publishing Group UK London}
}

@article{li2025cnn,
  title={{CNN}-Informer: A hybrid deep learning model for seizure detection on long-term {EEG}},
  author={Li, Chuanyu and Li, Haotian and Dong, Xingchen and Zhong, Xiangwen and Cui, Haozhou and Ji, Dezan and He, Landi and Liu, Guoyang and Zhou, Weidong},
  journal={Neural Networks},
  volume={181},
  pages={106855},
  year={2025},
  publisher={Elsevier}
}

@article{zhou2024lma,
  title={{LMA-EEGNet}: A lightweight multi-attention network for neonatal seizure detection using {EEG} signals},
  author={Zhou, Weicheng and Zheng, Wei and Feng, Youbing and Li, Xiaolong},
  journal={Electronics},
  volume={13},
  number={12},
  pages={2354},
  year={2024},
  publisher={MDPI}
}

@article{wang2024combining,
  title={Combining {EEG} features and convolutional autoencoder for neonatal seizure detection},
  author={Wang, Yuxia and Yuan, Shasha and Liu, Jin-Xing and Hu, Wenrong and Jia, Qingwei and Xu, Fangzhou},
  journal={International Journal of Neural Systems},
  volume={34},
  number={08},
  pages={2450040},
  year={2024},
  publisher={World Scientific}
}

@article{kim2023skin,
  title={Skin-interfaced wireless biosensors for perinatal and paediatric health},
  author={Kim, Joohee and Yoo, Seonggwang and Liu, Claire and Kwak, Sung Soo and Walter, Jessica R and Xu, Shuai and Rogers, John A},
  journal={Nature Reviews Bioengineering},
  volume={1},
  number={9},
  pages={631--647},
  year={2023},
  publisher={Nature Publishing Group UK London}
}

@article{kwak2021skin,
  title={Skin-integrated devices with soft, holey architectures for wireless physiological monitoring, with applications in the neonatal intensive care unit},
  author={Kwak, Sung Soo and Yoo, Seonggwang and Avila, Raudel and Chung, Ha Uk and Jeong, Hyoyoung and Liu, Claire and Vogl, Jamie L and Kim, Joohee and Yoon, Hong-Joon and Park, Yoonseok and others},
  journal={Advanced Materials},
  volume={33},
  number={44},
  pages={2103974},
  year={2021},
  publisher={Wiley Online Library}
}

\end{document}